\documentclass[epj]{svjour}
\usepackage{latexsym}
\usepackage{graphics}%
\usepackage{hyperref}%
\hypersetup{colorlinks = true, allcolors = blue}%
\usepackage[utf8]{inputenc}%
\usepackage{authblk}%
\usepackage{hepparticles}%
\usepackage{hepunits}%
\usepackage{hepnames}%

\newcommand{\rts}{\sqrt{s}}%
\newcommand{\epem}{\mathrm{e^+e^-}}%
\newcommand{\ffbar}{\Pf\overline{\Pf}}%
\newcommand{\bbbar}{\Pqb\overline{\Pqb}}%
\newcommand{\ccbar}{\Pqc\overline{\Pqc}}%
\newcommand{\mumu}{\mu^+\mu^-}%
\newcommand{\MZ}{m_{\PZ}}%
\newcommand{\MZSq}{m^2_{\PZ}}%
\newcommand{\GZ}{\Gamma_{\PZ}}%
\newcommand{\Gh}{\Gamma_{\rm had}}%
\newcommand{\Gb}{\Gamma_{\Pqb}}%
\newcommand{\Gc}{\Gamma_{\Pqc}}%
\newcommand{\Gl}{\Gamma_{\ell}}%
\newcommand{\Gn}{\Gamma_{\nu}}%
\newcommand{\Ginv}{\Gamma_{\rm inv}}%
\newcommand{\Rl}{R_{\ell}}%

\newcommand{\sinSqEff}{\mathrm{\sin^2\theta_W^{eff}}}%
\newcommand{\sinf}{\mathrm{\sin^2\theta_W^{\Pf,eff}}}%
\newcommand{\rhof}{\mathrm{\overline{\rho}_{\Pf}}}%

\newcommand{\Rb}{\ensuremath{R_{\rm b}}}%
\newcommand{\Rc}{\ensuremath{R_{\rm c}}}%
\newcommand{\AFBb}{\ensuremath{{A_{\rm FB}^{\rm b}}}}%
\newcommand{\AFBm}{\ensuremath{{A_{\rm FB}^{\mu}}}}%
\newcommand{\AFBf}{\ensuremath{{A_{\rm FB}^{\Pf}}}}%
\newcommand{\AFBzerob}{\ensuremath{{A_{\rm FB}^{0,{\rm b}}}}}%
\newcommand{\AFBc}{\ensuremath{{A_{\rm FB}^{\rm c}}}}%
\newcommand{\AFBzeroc}{\ensuremath{{A_{\rm FB}^{0,{\rm c}}}}}%
\newcommand{\Af}{\ensuremath{\mathcal{A}_{\Pf}}}%
\newcommand{\Ae}{\mathcal{A}_{\Pe}}%
\newcommand{\Atau}{\mathcal{A}_{\Ptau}}%
\newcommand{\Poltau}{\mathcal{P}_{\Ptau}}%
\newcommand{\AFBpol}{\ensuremath{{A_{\rm FB}^{\rm pol,\Ptau}}}}%

\newcommand{\alphas}{\ensuremath{\alpha_\mathrm{S}}}%
\newcommand{\alphasMZ}{\alphas(\MZSq)}%
\newcommand{\alphaMZ}{\alpha(\MZSq)}%
\newcommand{\eegg}{\ensuremath{\epem\to\gamma\gamma}}%

\begin{document}
\title{The Z lineshape challenge: ppm and keV measurements}
\author{Juan Alcaraz Maestre\inst{1} \and Alain Blondel\inst{2} \and Mogens Dam\inst{3} \and Patrick Janot\inst{4} 
}                     
\offprints{}          
\institute{CIEMAT, Madrid, Spain \and LPNHE, IN2P3/CNRS, Paris France, and University of Geneva, Switzerland \and Niels Bohr Institute, Copenhagen, Denmark \and CERN, EP Department, Geneva, Switzerland 
}
\date{
{\sl (Submitted to EPJ+ special issue: A future Higgs and Electroweak factory (FCC): Challenges towards discovery, Focus on FCC-ee)}
}
%
\abstract{
The FCC-ee offers powerful opportunities for direct or indirect evidence for physics beyond the Standard Model, via a combination of high precision measurements and searches for forbidden and rare processes and feebly coupled particles. A key element of FCC-ee physics program is the measurement of the Z lineshape from a total of $5\times 10^{12}$ Z bosons and a beam-energy calibration with relative uncertainty of $10^{-6}$. With this exceptionally large event sample, five orders of magnitude larger than that accumulated during the whole LEP1 operation at the Z pole, the defining parameters - $m_{\rm Z}$, $\Gamma_{\rm Z}$, $N_\nu$, $\sin^2\theta_{\rm W}^{\rm eff}$, $\alpha_{\rm S}(m_{\rm Z}^2)$, and $\alpha_{\rm QED}(m^2_{\rm Z})$ - can be extracted with a leap in accuracy of up to two orders of magnitude with respect to the current state of the art. The ultimate goal that experimental and theory systematic errors match the statistical accuracy (4\,keV on the Z mass and width, $3\times 10^{-6}$ on $\sin^2\theta_{\rm W}^{\rm eff}$, a relative $3\times 10^{-5}$ on $\alpha_{\rm QED}$, and less than 0.0001 on $\alpha_{\rm S}$) leads to highly demanding requirements on collider operation, beam instrumentation, detector design, computing facilities, theoretical calculations, and Monte Carlo event generators. Such precise measurements also call for innovative analysis methods, which require a joint effort and understanding between theorists, experimenters, and accelerator teams.\\
\PACS{
      {PACS-key}{describing text of that key}   \and
      {PACS-key}{describing text of that key}
     } 
} 
\maketitle

\section{Introduction}
\label{section:intro}

With an integrated luminosity of $150\,\mathrm{ab}^{-1}$ collected in $\approx 4$ years of running at centre-of-mass energies between 88 and 94\,GeV, FCC-ee~\cite{Benedikt:2651299} offers a unique opportunity to perform ultra-precise electroweak measurements of the Z resonance. More than $5\times 10^{12}$ Z decays, constituting the so-called TeraZ scenario, will be available for study. The statistical power of this sample is complemented by an extraordinary precision in the knowledge of the collision energy, of $\approx 100\,\mathrm{keV}$~\cite{R0polblondel}. Robust procedures to monitor other relevant beam collision parameters and the relative uncertainties between the energy points in Z lineshape scans~\cite{Blondel:2019jmp} are also an integral part of the physics program. 

The increase in luminosity with respect to past LEP experiments translates into an increase of more than two orders of magnitude in statistical sensitivity. With such large potential improvement over previous measurements, a key question is how much the associated systematic uncertainties can be reduced,  and ultimately match the statistical uncertainties. Challenges arise both at the theoretical and experimental levels. New theoretical paths will have to be pursued in order to provide most precise predictions for experimental observables~\cite{Blondel:2018mad}, and detailed experimental studies will have to be performed to optimize accelerator and detector designs. Exploring new analysis strategies and observables to simultaneously reduce both theoretical and experimental uncertainties will be another essential component of the challenge. 

An initial review of the FCC-ee potential regarding ultra-precise measurements at the Z pole can be found in Ref.~\cite{Abada:2019lih}. In the following we focus on some elements of the challenge that we consider relevant for success. The physics implications of the proposed program are expected to be deep. Regarding universal deviations, the new estimates of the oblique parameters $S$ and $T$ will constitute an early probe for Higgs compositeness or new interactions possibly occurring at the deca-TeV scale~\cite{Blondel:2021ema,Strategy:2019vxc}, i.e. one order of magnitude above currently explored scales. Deviations from universality related with these scales will be searched for through the precise measurement of the couplings of each fermion flavour to the Z. Separate access to the left- and right-handed components of the couplings will be available at FCC-ee even in the absence of polarised beams~\cite{Blondel:2019yqr}, as described below. 

\section{General considerations for a successful TeraZ program}

Regarding cross-section measurements, relative statistical uncertainties are expected in the $1/\sqrt{N}=10^{-5}$-$10^{-6}$ range, where $N$ is the number of events selected in the decay channel under study. Similar statistical uncertainties, of the order of $\scriptstyle{\sqrt{(1-\mathcal{A}^2)/N}}$, are expected on the absolute value of measured asymmetries $\mathcal{A}$.

At the theoretical front, and focusing on cross-section measurements, a limiting factor is the precision of the theoretical predictions used in luminosity measurements. Current studies using Bhabha scattering at low angle consider a relative accuracy of order $10^{-4}$ as a realistic target~\cite{Jadach:2018jjo}. The $\eegg$ process has also been suggested as an alternative channel to consolidate a $10^{-4}$ precision measurement, owing to the almost negligible size of the theoretical uncertainties of hadronic origin for this process~\cite{CarloniCalame:2019dom}. Cross-section ratio and asymmetry measurements are not affected by luminosity uncertainties. Besides the calculation of missing higher orders in these observables, reaching precisions of order $10^{-5}$ or better will probably require a deep change of philosophy in the measurement of lineshape parameters: inclusion of non-factorizable terms; interference effects between initial and final state radiation; redefinition of electroweak parameters at the amplitude level; fits of complete differential distributions using dedicated weighted Monte Carlo calculations, etc.~\cite{Blondel:2018mad} For the discussion below, and in order to better illustrate the main challenges, we nevertheless assume a simplified LEP-like strategy in the measurement of electroweak parameters~\cite{ALEPH:2005ab}. 

Focusing on experimental aspects, a typical limiting factor for cross-section measurements is the systematic uncertainty on the acceptance determination. A $10^{-5}$ uncertainty, even in processes presenting a relatively smooth behaviour of the 
angular distributions, implies a knowledge of the positions of the edges of sub-detectors at the $10\,\mu\mathrm{m}$ level over distances of the order of a meter. A first consequence is that detectors should be as homogeneous as possible. Such a precision is a realistic target given current tracking accuracy, but it demands dedicated efforts in terms of metrology, alignment, monitoring and designs able to ensure the stability of large detector volumes as a function of time. The challenge is even bigger for detectors located at very low polar angles and measuring differential cross sections with a $d\sigma/d\theta \propto 1/\sin\theta$ behaviour. For instance, a luminosity monitor located at 1\,m of the interaction point with an inner radius of $\approx 65$\,mm demands a $1\,\mu\mathrm{m}$ ($1\,\mu \mathrm{rad}$) precision in positioning, in order to reach $10^{-4}$ uncertainties~\cite{Benedikt:2651299}. Other requirements imposed by acceptance systematics are the uniformity in the detector response, redundant particle identification capabilities, beam stability and a detailed monitoring of the beam geometry conditions at the interaction point. 

\section{Z lineshape determination and ppm/keV precision observables}
\label{section:xsections}

At FCC-ee Z lineshape scans are expected to provide a measurement of $\MZ$ with unprecedented precision, $\delta\MZ \approx 0.1\,\mathrm{MeV}$, i.e., $\approx 20$ times better than the present precision from LEP. The uncertainty is fully dominated by the uncertainty on the collision energy, which can be determined with $\approx 100\,\mathrm{keV}$ precision using resonant depolarisation of the transversely polarised beams~\cite{Blondel:2019jmp}. This method was already used at LEP and will be significantly improved at FCC-ee. There will be very frequent in-situ calibrations using non-colliding  pilot bunches, simultaneous with nominal collisions. Nevertheless, detailed studies of the differences between colliding and pilot bunches will be necessary. Also, the beam energy spread and the energy asymmetries between the two beams should be monitored via analysis of the longitudinal boosts of $\epem\to\mumu$ events~\cite{Blondel:2019jmp}. 

The total Z width, $\GZ$, is directly connected with the width of the Z lineshape. A statistical precision of $4\,\mathrm{keV}$ is expected from a fit of the hadronic lineshape. The overall precision is dominated by the so-called ``point-to-point'' uncertainties, which correspond to systematic differences between their central $\rts$ values that are not 100\% positively correlated. The high statistics and the expected muon momentum resolution of tracking detectors ($\delta \left( 1/p_{\rm T} \right) \approx$ a few $10^{-5}$ for $p = 45$\,GeV at normal incidence) allow a quantification of these differences from the dimuon invariant mass distributions obtained at each collision energy. An uncertainty of $\approx 25\,\mathrm{keV}$ was obtained in preliminary studies~\cite{Blondel:2019jmp}, which translates into a precision improvement of two orders of magnitude over current LEP results. 
re
The ratio between hadronic and leptonic cross sections, $\Rl$, is an essential observable for the extraction from the total width of the global and individual leptonic and hadronic partial widths. The global leptonic partial width is a direct test of new physics involving weak isospin violations, while the individual partial widths constitute a powerful test of lepton coupling universality in the neutral current. $\Rl$ also provides one of the most precise ways to measure $\alphasMZ$~\cite{Gomez-Ceballos:2013zzn,Benedikt:2651299}. Its measurement is independent of luminosity uncertainties, and therefore relative precisions below $10^{-4}$ can be contemplated. Improvements over LEP results by a factor ranging between $20$ and $100$, i.e., $\delta\Rl/\Rl = (1-5)\times 10^{-5}$ are expected. Leptonic and hadronic Z decays provide clean signatures at the Z pole, and are only affected by limited backgrounds at small visible mass (two-photon processes). According to LEP experience, the largest source of uncertainty could be the acceptance of the leptonic channels. Even with a rather hermetic detector and sufficiently redundant identification criteria, the edges of the tracking acceptance and the interplay with beam position and width parameters will have to be understood in detail. A precision in the position of these edges at the level of of $10\,\mu\mathrm{m}$ might be required. 

A measurement of the relative invisible width $\Ginv/\Gl$, or equivalently of the number of neutrinos, $N_{\nu}= \Ginv/\Gn$, involves the measurement of the Z peak cross section~\cite{ALEPH:2005ab,Voutsinas:2019hwu,Janot:2019oyi}. The measurement is therefore limited by the precision in the measurement of the luminosity, as commented before. A $10^{-4}$ precision represents already one order of magnitude improvement with respect to LEP for a similar luminosity detector coverage. The study of radiative recoil ratios like $\sigma(\Pnu\APnu\gamma)/\sigma(\ell^+\ell^-\gamma)$ above the Z pole is also being considered as an alternative at a similar level of precision, dominated by statistical uncertainties~\cite{Abada:2019lih}. 

\section{Precise measurements of the electroweak fermion couplings to the Z and of \texorpdfstring{$\alphaMZ$}{qed}}
\label{section:asymmetries}

One of the main targets of the electroweak program is the precise measurement of the chiral couplings to the Z for each individual fermion. The baseline FCC-ee proposal relies on a direct measurement of the Z polarisation induced by the beam particles with studies of the $\tau$ polarisation as a function of the polar angle. This approach suppresses the need of longitudinal beam polarisation, which would otherwise imply a non-negligible loss in luminosity. 
The tau polarisation in Z decay is measured from the charged particle momentum distribution in the semi-leptonic decays $\tau \to {\rm e} \nu_{\rm e} \nu_\tau$, $\mu \nu_\mu \nu_\tau$ or in hadronic decays $\tau \rightarrow {\rm h} \nu_\tau$ where  $\rm h$ can be $\pi$, ${\rm K}$,  $\rho$, ${\rm K}^\ast$, ${\rm a}_1$, etc. Each channel has a different polarisation analysis power and must therefore be analysed independently. A clean separation between channels is also essential. The analysis of the $\tau$ polarisation dependence on the $\epem\to\tau^+\tau^-$ scattering angle $\theta$ gives access to both the tau and electron chiral coupling asymmetries ${\cal A}_\tau$ and ${\cal A}_{\rm e}$ independently
\begin{equation}
P(\cos\theta)   =  - \frac{  {\cal A}_\tau ( 1 + \cos^2 \theta)  + 2 {\cal A}_{\rm e} \cos\theta    } {  (1 + \cos^2 \theta) + 2 {\cal A}_{\rm e} {\cal A}_\tau \cos\theta  }, 
\end{equation}
%
The previous expression is valid at Born-improved level and in the massless lepton limit. $\Af$ is the chiral asymmetry parameter of the fermion $\Pf$ in its coupling to the Z. 
The average $\tau$ polarisation, $\Poltau$, provides a direct measurement of $\Atau$: $\Poltau=-\Atau$, whereas the forward-backward polarisation asymmetry, $\AFBpol$, provides a direct measurement of $\Ae$, the induced Z polarisation in the $\epem$ collision: $\AFBpol=-\frac{3}{4}\Ae$. The remaining systematic uncertainty on $\Ae$ at LEP was originating from the limited knowledge of non-$\tau$ backgrounds~\cite{Heister:2001uh}. At FCC-ee, huge control samples will be available to reduce this component, but dedicated studies are still necessary to estimate the ultimately reachable precision.

The forward-backward asymmetry in the $\epem\to\ffbar$ process, $\AFBf = \frac{3}{4}\Ae\Af$ gives access to the chiral couplings of the fermion $\Pf$ when combined with tau polarisation studies. Chiral couplings can be trivially converted in measurements of vector and axial couplings or, alternatively, in a Born-improved spirit, in measurements of the effective $\rhof$ and $\sinf$ parameters for each individual fermion. Assuming universal deviations, $\sinf$ measurements become a measurement of the effective weak mixing angle $\sinSqEff$. The muon channel is particularly promising in this respect, aiming for a precision approaching the ultimate statistical sensitivity of $3\times 10^{-6}$ on $\AFBm$. At that level of precision, further theoretical studies of QED corrections and in particular of interference effects between initial and final-state radiation will be mandatory. Detailed studies of the beam parameters and of the polar angular resolution will be required as well~\cite{Blondel:2019jmp,Jadach:2018lwm}. 

The extreme FCC-ee precision requirements demand an equivalent level of precision on the input parameters to theoretical predictions. One of these parameters is $\alphaMZ$, the electromagnetic coupling constant at the Z scale, which is the source of one of the dominant uncertainties in present fits. In practice, $\alphaMZ$ can be treated as another parameter to be determined in the Z lineshape running. This is supported by the study presented in Ref.~\cite{Janot:2015gjr}, which proposes a measurement of $\alphaMZ$ with a $3\times 10^{-5}$ relative precision, largely uncorrelated with other Z lineshape parameters. Most of the sensitivity lies in the linear dependence of the $\AFBm$ asymmetry with respect to the $\Pgamma-\PZ$ interference term around the Z peak. An appropriate combination of measurements at the energy points with the largest sensitivity, $\sqrt{s}=87.9$ and $94.3$~GeV, using one year of integrated luminosity should provide the required precision. Still, reaching the aimed precision will require the calculation of missing electroweak corrections of higher order, as well as more detailed studies of initial-final state interference effects. 

\section{Heavy quark precision measurements}
\label{section:hq}

The cross-section ratios and asymmetries $\Rb\equiv \Gb/\Gh$, $\Rc\equiv \Gc/\Gh$, $\AFBb$, and $\AFBc$ are expected to be measured at FCC-ee with more than one order of magnitude better precision than at LEP/SLC. On top of the increase in the number of collected events, the statistical precision should be boosted with respect to LEP/SLC by the much higher b- or c-tagging efficiencies and purities obtained with better detectors and more powerful lifetime tagging techniques. Systematic uncertainties are expected to be significantly reduced by the use of exclusive B decays and enriched control samples in specific regions of phase space. 

At FCC-ee, $\Rb$ and $\Rc$ will be likely measured with double-tagging techniques on both hemispheres of a $\bbbar$ or a $\ccbar$ event~\cite{ALEPH:2005ab}. This strategy allows a measurement independent of the knowledge of the tagging efficiency in the limit of negligible backgrounds and correlations between  hemispheres. At LEP, hemisphere correlations due to QCD effects (mostly hard gluon emission) and primary vertex determination were dominant sources of uncertainty. Gluon splitting increases the number of single tags in events with two light-flavour jets, and also constitutes a significant source of correlated uncertainty between experiments. The vertex precision of the new generation of detectors should contribute to reduce vertex correlations significantly.  Studies as a function of the acoplanarity between b-tagged jets can help reducing QCD correlations, and huge gluon splitting samples will become available for a precise understanding of this source of uncertainty. Nevertheless, coming studies are needed to quantify these improvements in more detail.

The bare forward-backward asymmetry of b quarks at the Z pole, $\AFBzerob$, is the electroweak observable that currently presents the largest deviation with respect to the standard model expectation in current fits~\cite{ALEPH:2005ab} ($\approx 3\sigma$ pull). An order-of-magnitude improved measurement at FCC-ee could thus become a clean signal of new physics if the deviation in the central value is confirmed. The world-average measurement is still dominated by statistical uncertainties ($\delta\AFBzerob({\rm stat.}) = 0.0016$), but is also affected by non-negligible systematic uncertainties ($\delta\AFBzerob({\rm syst.})=0.0007$). A fraction of it can be reduced at FCC-ee through dedicated studies on high-statistics control samples. A detailed analysis of the detector requirements to maximise flavour identification capabilities is also mandatory. Exclusive B decays can be exploited as well. For instance, about $10^8$ $\mathrm{B}^+$ decays, not affected by charm contamination or B-mixing effects, will be available at FCC-ee~\cite{Fabrizio-afbb}. The expected improvements in flavour tagging will be much more visible in the case of $\AFBzeroc$ measurements. In addition, the availability of exclusive, high purity, and large statistics D decay samples should provide a significant improvement in terms of precision compared with LEP measurements~\cite{ALEPH:2005ab}. There, most measurements were performed with inclusive or pseudo-inclusive techniques on samples with significant b-quark contamination. Dedicated studies are certainly necessary. 

An irreducible source of uncertainty in the current estimate of $\AFBzerob$, fully correlated among experiments, is the presence of a QCD correction factor, of order $1-\alphas/\pi$, that accounts for the shift between the experimentally observed asymmetry and $\AFBzerob$~\cite{Djouadi-afbb,Altarelli:1992fs,ep-98}. The main role of the correction is to absorb the angular distortions due to final-state QCD radiation in the $\mathrm{Z}\to\mathrm{b\overline{b}(g)}$ decay. Recent re-evaluations of that uncertainty~\cite{dEnterria:2018jsx} based on modern parton shower tunes seem to be consistent with the initial estimates. New strategies to reduce or constrain experimentally the size of these uncertainties are being developed, for an initial target of $\delta\AFBzerob \approx 0.0001$~\cite{AlcarazMaestre:2020fmp}.

\section{Outlook}
\label{section:outlook}
Besides theory requirements, discussed in more detail in~\cite{Freitas:2019bre,Heinemeyer:2021rgq}, careful experimental studies with realistic detector descriptions are necessary to estimate the ultimate precision for FCC-ee electroweak measurements at the Z pole. In the particular case of cross section measurements, aiming for a precision of $\approx 10^{-4}$ is already imposing severe constraints on the design and tolerances of luminosity monitors. Regarding final states involving heavy flavours (tau, bottom, charm), an improvement of one order of magnitude with respect to previous LEP/SLC measurements may imply new constraints on the detector design and the development of new, more powerful tagging techniques. Let us note that the huge available statistics at the Z peak suggests that exclusive decay identification should be explored in more depth, as a complementary path to reduce systematic uncertainties in some particular cases. Addressing all these challenges is a critical step for success and, as such, one of the main objectives of the present electroweak physics program at FCC-ee.


%
%
%

\bibliographystyle{myutphys}
\bibliography{references}

\providecommand{\href}[2]{#2}\begingroup\raggedright\begin{thebibliography}{10}

\bibitem{Benedikt:2651299}
A.~Abada {\em et~al.}, ``{FCC-ee: The Lepton Collider}; {FCC CDR Vol. 2}'',
  \href{http://dx.doi.org/10.1140/epjst/e2019-900045-4}{{\em Eur. Phys. J. ST}
  {\bfseries 228} no.~2, (2019) 261--623}.

\bibitem{R0polblondel}
A.~Blondel and E.~Gianfelice, ``The challenges of beam polarization and
  kev-scale centre-of-mass energy calibration.'' {A future Higgs and
  Electroweak factory (FCC): Challenges towards discovery, EPJ+ special issue,
  Focus on FCC-ee}.

\bibitem{Blondel:2019jmp}
A.~Blondel, P.~Janot, J.~Wenninger, {\em et~al.}, ``{Polarization and
  Centre-of-mass Energy Calibration at FCC-ee}'',
\href{http://arxiv.org/abs/1909.12245}{{\ttfamily arXiv:1909.12245
  [physics.acc-ph]}}.

\bibitem{Blondel:2018mad}
A.~Blondel {\em et~al.},
  \href{http://dx.doi.org/10.23731/CYRM-2019-003}{``{Standard model theory for
  the FCC-ee Tera-Z stage}'',} in {\em {Mini Workshop on Precision EW and QCD
  Calculations for the FCC Studies : Methods and Techniques}}, vol.~3/2019 of
  {\em CERN Yellow Reports: Monographs}.
\newblock CERN, Geneva, 9, 2018.
\newblock \href{http://arxiv.org/abs/1809.01830}{{\ttfamily arXiv:1809.01830
  [hep-ph]}}.

\bibitem{Abada:2019lih}
A.~Abada {\em et~al.}, ``{FCC Physics Opportunities}; {FCC CDR Vol. 1}'',
  \href{http://dx.doi.org/10.1140/epjc/s10052-019-6904-3}{{\em Eur. Phys. J. C}
  {\bfseries 79} no.~6, (2019) 474}.

\bibitem{Blondel:2021ema}
A.~Blondel and P.~Janot, ``{FCC-ee overview: new opportunities create new
  challenges, {\rm in A future Higgs and Electroweak factory (FCC): Challenges
  towards discovery, EPJ+ special issue, Focus on FCC-ee}}'',
  \href{http://arxiv.org/abs/2106.13885}{{\ttfamily arXiv:2106.13885
  [hep-ex]}}.

\bibitem{Strategy:2019vxc}
R.~K. Ellis {\em et~al.}, ``{Physics Briefing Book}: {Input for the European
  Strategy for Particle Physics Update 2020}'',
  \href{http://arxiv.org/abs/1910.11775}{{\ttfamily arXiv:1910.11775
  [hep-ex]}}.

\bibitem{Blondel:2019yqr}
N.~Alipour~Tehrani {\em et~al.}, ``{FCC-ee: Your Questions Answered, {\rm
  Section 8.2}}'', in {\em {CERN Council Open Symposium on the Update of
  European Strategy for Particle Physics (EPPSU) Granada, Spain, May 13-16,
  2019}}, A.~Blondel and P.~Janot, eds.
\newblock 2019.
\newblock
\href{http://arxiv.org/abs/1906.02693}{{\ttfamily arXiv:1906.02693 [hep-ph]}}.
\newblock

\bibitem{Jadach:2018jjo}
S.~Jadach, W.~P\l{}aczek, M.~Skrzypek, B.~F.~L. Ward, and S.~A. Yost, ``{The
  path to 0.01\% theoretical luminosity precision for the FCC-ee}'',
  \href{http://dx.doi.org/10.1016/j.physletb.2019.01.012}{{\em Phys. Lett. B}
  {\bfseries 790} (2019) 314--321},
  \href{http://arxiv.org/abs/1812.01004}{{\ttfamily arXiv:1812.01004
  [hep-ph]}}.

\bibitem{CarloniCalame:2019dom}
C.~M. Carloni~Calame, M.~Chiesa, G.~Montagna, O.~Nicrosini, and F.~Piccinini,
  ``{Electroweak corrections to $e^+e^-\to\gamma\gamma$ as a luminosity process
  at FCC-ee}'', \href{http://dx.doi.org/10.1016/j.physletb.2019.134976}{{\em
  Phys. Lett. B} {\bfseries 798} (2019) 134976},
  \href{http://arxiv.org/abs/1906.08056}{{\ttfamily arXiv:1906.08056
  [hep-ph]}}.

\bibitem{ALEPH:2005ab}
{ALEPH, DELPHI, L3, OPAL, SLD, LEP Electroweak Working Group, SLD Electroweak
  Group, SLD Heavy Flavour Group}, S.~Schael {\em et~al.}, ``{Precision
  electroweak measurements on the $Z$ resonance}'',
  \href{http://dx.doi.org/10.1016/j.physrep.2005.12.006}{{\em Phys. Rept.}
  {\bfseries 427} (2006) 257--454},
  \href{http://arxiv.org/abs/hep-ex/0509008}{{\ttfamily arXiv:hep-ex/0509008}}.

\bibitem{Gomez-Ceballos:2013zzn}
{TLEP Design Study Working Group}, M.~Bicer {\em et~al.}, ``{First Look at the
  Physics Case of TLEP}'',
  \href{http://dx.doi.org/10.1007/JHEP01(2014)164}{{\em JHEP} {\bfseries 01}
  (2014) 164},
\href{http://arxiv.org/abs/1308.6176}{{\ttfamily arXiv:1308.6176 [hep-ex]}}.

\bibitem{Voutsinas:2019hwu}
G.~Voutsinas, E.~Perez, M.~Dam, and P.~Janot, ``{Beam-beam effects on the
  luminosity measurement at LEP and the number of light neutrino species}'',
  \href{http://dx.doi.org/10.1016/j.physletb.2019.135068}{{\em Phys. Lett. B}
  {\bfseries 800} (2020) 135068},
  \href{http://arxiv.org/abs/1908.01704}{{\ttfamily arXiv:1908.01704
  [hep-ex]}}.

\bibitem{Janot:2019oyi}
P.~Janot and S.~Jadach, ``{Improved Bhabha cross section at LEP and the number
  of light neutrino species}'',
  \href{http://dx.doi.org/10.1016/j.physletb.2020.135319}{{\em Phys. Lett. B}
  {\bfseries 803} (2020) 135319},
  \href{http://arxiv.org/abs/1912.02067}{{\ttfamily arXiv:1912.02067
  [hep-ph]}}.

\bibitem{Heister:2001uh}
{ALEPH}, A.~Heister {\em et~al.}, ``{Measurement of the tau polarization at
  LEP}'', \href{http://dx.doi.org/10.1007/s100520100689}{{\em Eur. Phys. J. C}
  {\bfseries 20} (2001) 401--430},
  \href{http://arxiv.org/abs/hep-ex/0104038}{{\ttfamily arXiv:hep-ex/0104038}}.

\bibitem{Jadach:2018lwm}
S.~Jadach and S.~Yost, ``{QED Interference in Charge Asymmetry Near the Z
  Resonance at Future Electron-Positron Colliders}'',
  \href{http://dx.doi.org/10.1103/PhysRevD.100.013002}{{\em Phys. Rev. D}
  {\bfseries 100} no.~1, (2019) 013002},
  \href{http://arxiv.org/abs/1801.08611}{{\ttfamily arXiv:1801.08611
  [hep-ph]}}.

\bibitem{Janot:2015gjr}
P.~Janot, ``{Direct measurement of $\alpha_{QED}(m_{Z}^{2})$ at the FCC-ee}'',
  \href{http://dx.doi.org/10.1007/JHEP02(2016)053}{{\em JHEP} {\bfseries 02}
  (2016) 053}, \href{http://arxiv.org/abs/1512.05544}{{\ttfamily
  arXiv:1512.05544 [hep-ph]}}. [Erratum: JHEP 11, 164 (2017)].

\bibitem{Fabrizio-afbb}
F.~Palla, ``{Challenges for EW b physics measurements}.'' talk at the FCC Week
  2019, Brussels,
  \url{https://indico.cern.ch/event/727555/contributions/3446561/}, 2019.

\bibitem{Djouadi-afbb}
A.~Djouadi, B.~Lampe, and P.~Zerwas, ``{A Note on the QCD corrections to
  forward - backward asymmetries of heavy quark jets in Z decays}'',
  \href{http://dx.doi.org/10.1007/BF01564827}{{\em Z. Phys. C} {\bfseries 67}
  (1995) 123--128}, \href{http://arxiv.org/abs/hep-ph/9411386}{{\ttfamily
  arXiv:hep-ph/9411386}}.

\bibitem{Altarelli:1992fs}
G.~Altarelli and B.~Lampe, ``{Second order QCD corrections to heavy quark
  forward - backward asymmetries}'',
  \href{http://dx.doi.org/10.1016/0550-3213(93)90138-F}{{\em Nucl. Phys. B}
  {\bfseries 391} (1993) 3--22}.

\bibitem{ep-98}
{LEP Heavy Flavor Working Group}, D.~Abbaneo, P.~Antilogus, T.~Behnke,
  S.~Blyth, M.~Elsing, R.~Faccini, R.~Jones, K.~Monig, S.~Petzold, and
  R.~Tenchini, ``{QCD corrections to the forward - backward asymmetries of c
  and b quarks at the Z pole}'',
  \href{http://dx.doi.org/10.1007/s100520050196}{{\em Eur. Phys. J. C}
  {\bfseries 4} (1998) 185--191}.

\bibitem{dEnterria:2018jsx}
D.~d'Enterria and C.~Yan, ``{Forward-backward $b$-quark asymmetry at the Z
  pole: QCD uncertainties redux}'', in {\em {Proceedings, 53rd Rencontres de
  Moriond on QCD and High Energy Interactions (Moriond QCD 2018): La Thuile,
  Italy, March 17-24, 2018}}, pp.~253--257.
\newblock 2018.
\newblock
\href{http://arxiv.org/abs/1806.00141}{{\ttfamily arXiv:1806.00141 [hep-ex]}}.
\newblock

\bibitem{AlcarazMaestre:2020fmp}
J.~Alcaraz~Maestre, ``{Revisiting QCD corrections to the forward-backward
  charge asymmetry of heavy quarks in electron-positron collisions at the Z
  pole: really a problem?}'', \href{http://arxiv.org/abs/2010.08604}{{\ttfamily
  arXiv:2010.08604 [hep-ph]}}.

\bibitem{Freitas:2019bre}
A.~Freitas {\em et~al.}, ``{Theoretical uncertainties for electroweak and
  Higgs-boson precision measurements at FCC-ee}'',
  \href{http://arxiv.org/abs/1906.05379}{{\ttfamily arXiv:1906.05379
  [hep-ph]}}.

\bibitem{Heinemeyer:2021rgq}
S.~Heinemeyer, S.~Jadach, and J.~Reuter, ``{Theory requirements for SM Higgs
  and EW precision physics at the FCC-ee, {\rm in A future Higgs and
  Electroweak factory (FCC): Challenges towards discovery, EPJ+ special issue,
  Focus on FCC-ee}}'', \href{http://arxiv.org/abs/2106.11802}{{\ttfamily
  arXiv:2106.11802 [hep-ph]}}.

\end{thebibliography}\endgroup
\end{document}